\documentclass[mathleft
]{an}
\usepackage{graphicx}
\usepackage{times}
\begin{document}

\Pagespan{789}{}
\Yearpublication{2006}
\Yearsubmission{2006}
\Month{9}  
\Volume{327}  
\Issue{9}
\DOI{10.1002/asna.2006}

\title{Remarks on statistical errors in equivalent widths}

\author{K. Vollmann \and T. Eversberg\thanks{Corresponding author:
  \email{eversberg@stsci.de}\newline}}

\titlerunning{Statistical errors in equivalent widths}
\authorrunning{K. Vollmann \& T. Eversberg}
\institute{Schn\"orringen Telescope Science Institute, Ringweg 7, 
51545 Waldbr\"ol, Germany, www.stsci.de}

\received{2006 Mar 29}    \accepted{xxxxx}   
\publonline{xxxxx}

\abstract{Equivalent width measurements for rapid line variability in atomic 
spectral lines are degraded by increasing error bars with shorter exposure 
times. We derive an expression for the error of the line equivalent width 
$\sigma(W_\lambda)$ with respect to pure photon noise statistics and provide 
a correction value for previous calculations.}

\maketitle

\section{Introduction}
Error bars of line equivalent widths $W_{\lambda}$ have been the subject of a 
number of investigations of high temporal and spectral resolution observations 
with exposures of  the order of minutes or even seconds, leading to large 
errors. A first expression for the calculation of $\sigma(W_{\lambda})$ for 
pure photon noise statistics was estimated by Williams et al. (\cite{wil74}) 
for their rapid line variability observations of Ap and Bp stars. This was 
then supplemented by Young (\cite{you74}) for  errors due to scintillation. Lacy 
(\cite{lac77}) published an improvement of Williams et al. and compared pure 
photon noise statistics with the case including scintillation for his 
$H\alpha$ and $H\beta$ rapid line investigations of massive stars. Guided by 
his work Chalabaev \& Maillard (\cite{cha83}) again derived an expression for 
the error from pure photon noise statistics in an appendix of their paper on 
rapid spectral variability of Be stars. However, their derivation of 
$\sigma(W_{\lambda})$ shows some inconsistencies which we attempt to resolve in 
this paper. 

\section{The equivalent width of a spectroscopic line}

In practice, the measurement of noise within a line can be a difficult task, 
because photon noise is superposed with  stellar variations.
Hence, it is necessary to estimate an expression for the equivalent width 
which separates these two parameters and then to find an expression for its 
error $\sigma(W_{\lambda})$. We start with the standard definition of the 
equivalent width:
	\[
W_{\lambda} = \int_{\lambda_1}^{\lambda_2} \frac{F_c(\lambda) - 
F(\lambda)}{F_c(\lambda)} d\lambda
\]
or
\begin{equation}
W_{\lambda} = \int_{\lambda_1}^{\lambda_2} \left[1-\frac{F(\lambda)}{F_c(\lambda)}
\right] d\lambda
\label{eq1}
\end{equation}
with $F_c(\lambda)$ the flux in the continuum, $F(\lambda)$ the flux in the line at the wavelength $\lambda$ and
$F(\lambda) = F_c(\lambda)$ for $\lambda \ge \lambda_2$ and $\lambda \le 
\lambda_1$. In a first step we integrate  eq.\,(\ref{eq1}) separately and we 
obtain
\[
W_{\lambda} = \Delta\lambda - \int_{\lambda_1}^{\lambda_2} 
\frac{F(\lambda)}{F_c(\lambda)} d\lambda. 
\]
with the wavelength interval $\Delta\lambda = \lambda_2 - \lambda_1$. By 
applying the mean value theorem we obtain
\begin{equation}
W_{\lambda} = \Delta\lambda \cdot \left[ 1 - \overline{\left( \frac{F(\lambda)}
{F_c(\lambda)}\right)} \right]
\label{eq2}
\end{equation}
with $\overline{\left(\frac{F(\lambda)}{F_c(\lambda)}\right)}$ the normalized 
average flux within $\Delta\lambda$. Equation\,(\ref{eq2}) can be written 
as
\begin{equation}
W_{\lambda} \approx \Delta\lambda \cdot \left[ 1 - \frac{\overline{F}} {\overline{F_c}} 
\right].
\label{eq3}
\end{equation}

This can be shown by applying a discrete approximation of eq.\,(\ref{eq3}). We use the arithmetic mean 
and substitute the line and continuum fluxes by their average values plus 
their deviations $\Delta F_i$ and $\Delta F_{c_i}$. 
\begin{eqnarray}
\nonumber
\overline{\left(\frac{F}{F_c}\right)} = \frac{1}{M} \cdot 
\sum_{i=1}^M{\frac{F_i}{F_{c_i}}}
\Leftrightarrow
\overline{\left(\frac{F}{F_c}\right)} = \frac{1}{M} \cdot \sum_{i=1}^M \left( 
\frac{\overline{F} + \Delta F_i}{\overline{F_c} + \Delta F_{c_i}}\right)
\end{eqnarray}
$M$ represents the number of pixels within the line. Note that $\Delta F_i$ 
within the line consists not only of noise but also of line information and 
can be quite large in contrast to $\Delta F_{c_i}$. 
\[
\overline{\left(\frac{F}{F_c}\right)} = \frac{\overline{F}}{\overline{F_c}} 
\cdot \frac{1}{M} \cdot \sum_{i=1}^M{\frac{ (1+\frac{\Delta F_i}{\overline{F}})}{ (1+\frac{\Delta F_{c_i}}{\overline{F_c}})}}
\]In case of sufficient S/N within the continuum we have
\[
\left|\frac{\Delta F_{c_i}}{\overline{F_c}}\right| << 1
\Rightarrow
1+\frac{\Delta F_{c_i}}{\overline{F_c}} \approx 1 
\]
\[
\overline{\left(\frac{F}{F_c}\right)} = \frac{\overline{F}}{\overline{F_c}} \cdot 
\frac{1}{M} \cdot \sum_{i=1}^M{ (1+\frac{\Delta F_i}{\overline{F}})}
\]
\[
\overline{\left(\frac{F}{F_c}\right)} = \frac{\overline{F}}{\overline{F_c}} + 
\frac{1}{\overline{F_c}}  \cdot \frac{1}{M} \cdot \sum_{i=1}^M{ \Delta F_i}
\]
However, by definition
\[
 \frac{1}{M} \cdot \sum_{i=1}^M{ \Delta F_i} = 0 
\Longrightarrow \overline{\left(\frac{F}{F_c}\right)} \approx  
\frac{\overline{F}}{\overline{F_c}}. 
\]

Hence, the equivalent width can be estimated from the 
wavelength interval and the average intensities in the line and the continuum.

\section{The error of the equivalent width}
The intensities $F(\lambda)$ and $F_c(\lambda)$ are influenced by certain 
statistical errors. If we keep in mind that $F_c$ is generally estimated 
outside the line region and interpolated across the line where the line flux 
is measured and if we assume that the errors are not correlated we are able to 
estimate their standard deviations separately. 
Following the principal of error propagation we expand eq.\,(\ref{eq3}) in a 
Taylor series
	\[
W_{\lambda} = W(\overline{F},\overline{F_c}) + \frac{\partial W_{\lambda}}{\partial 
\overline{F}}(F-\overline{F})
+ \frac{\partial W_{\lambda}}{\partial \overline{F_c}}(F_c-\overline{F_c})
\]
where $F$ and $F_c$ are random variables. 

The variance $Var (W_{\lambda}) \equiv \sigma^2(W_{\lambda})$ of the expansion 
is
\begin{equation}
\sigma^2(W_{\lambda}) = \left[ \frac{\partial W_{\lambda}}{\partial \overline{F}} 
\cdot \sigma (F) \right]^2 + \left[ \frac{\partial W}{\partial \overline{F_c}} \cdot 
\sigma (F_c) \right]^2 
\label{eq10}
\end{equation}
with $\sigma (F)$ and $\sigma (F_c)$ the standard deviations in the line and 
the continuum, respectively. By using eq.\,(\ref{eq3}) the two partial 
derivatives are
	\[
\frac{\partial W_{\lambda}}{\partial \overline{F}} = - 
\frac{\Delta\lambda}{\overline{F_c}}
\]
with $\Delta\lambda = M\cdot h_{\lambda}$ and the spectral dispersion 
$h_{\lambda}$ and
	\[
\frac{\partial W_{\lambda}}{\partial \overline{F_c}} = \frac{1}{\overline{F_c}} 
\left(\Delta \lambda - W_{\lambda} \right).
\]

\subsection{Weak lines}

For weak lines the depth of the line is negligible and 
	\[
\sigma(F) = \frac{\overline{F}}{S/N} \cong \sigma(F_c) = \frac{\overline{F_c}}{S/N} 
\]
with the signal-to-noise ratio S/N in the undisturbed continuum. With these 
terms we can write 
\begin{eqnarray}
\sigma^2(W_{\lambda}) = \left[ \frac{M\cdot h_{\lambda}}{S/N} \right]^2 \cdot 
\left[\frac{\overline{F}}{\overline{F_c}}\right]^2 +\cr
+ \left[ \frac{\sigma(F_c)}{\overline{F_c}} \cdot (\Delta \lambda - W_{\lambda})\right]^2
\label{eq6}
\end{eqnarray}
where the first term is the photometric uncertainty and the second term the 
uncertainty of the continiuum estimation over the line. According to 
eq.\,(\ref{eq3}) both terms are identical and we have
\begin{equation}
\sigma(W_{\lambda}) = \sqrt{2} \cdot \frac{(\Delta\lambda - W_{\lambda})}{S/N}.
\label{eq8} 
\end{equation}

This result is easy to understand if we realize that the line strength for 
weak lines will be of the order of the noise itself.

\subsection{Low- and high-flux lines}
In the case of low (absorption) and high (emission) flux lines we have to use 
the corresponding poisson-statistic
	\[
\sigma(F)  = \sqrt{\frac{\overline{F}}{\overline{F_c}}} \cdot \sigma(F_c)
\]
and we get
\begin{equation}
\sigma(W_{\lambda}) =  \sqrt{1 + \frac{\overline{F_c}}{\overline{F}}} \cdot 
\frac{(\Delta\lambda - W_{\lambda})}{S/N}.
\label{eq4}
\end{equation}

This result seems to be quite simple. 
However, we should not expect a complex solution for the error bars of 
equivalent widths because of a relatively simple definition of $W$.
In both cases we now can obtain the standard deviations with the measurable 
parameters $S/N$, $\Delta\lambda$, $\overline{F}$ and $\overline{F_c}$. In 
addition, eq.\,(\ref{eq4}) represents the general error of the line equivalent 
width and for the case $\overline{F} \approx \overline{F_c}$ we obtain the 
result for weak lines again.

\section{A comparison with former calculations}
Interestingly our result in eq.\,(\ref{eq6}) differs from eq.\,(A9) of 
Chalabaev \& Maillard (C\&M) by a factor $M$, the number of pixels corresponding 
to the interval $\lambda_1 \le \lambda \le \lambda_2$, in the first summand. 
The reason for this difference can be found in their expansion C\&M\,(A3):

\begin{itemize}
	\item C\&M\,(A3) is a multidimensional expansion for the equivalent width with average values $\overline{F_j}$  for all pixel fluxes $F_j$ (the pixel fluxes $F_j$ define random variables). However, it is not clear why the expansion over $M$ fluxes $F_j$ should be necessary.   
	\item Although there is no principal difference between the line and the continuum the continuum part is not consequently developed as it is done for the line part. 
They also say that "the value (S/N) characterizes the spread of read-outs on the continuum of an individual spectrum, while the quantity $\sigma(\overline{F_c})$ is the total uncertainty of the continuum level determination, and these two quantities may be related in a non-simple way". However, we assume that the uncertainty of the continuum is simply defined by the noise within the continuum and therefore the above distinction is not necessary.

	\item Another difference between C\&M and our approach can be found in their eq.\,(A7) which gives $S/N = \frac{F_j}{\sigma(F_j)}$. However, the correct expression is $S/N = \frac{\overline{F}}{\sigma(F)}$ 
(which is $\frac{\overline{F_j}}{\sigma(F_j)}$ in C\&M notation).
\end{itemize}

\subsection{The $\sqrt{N}$-rule}
Generally, for the variance of a random variable $x$ we have
$
Var(x) = \overline{x^2} - \overline{x}^2
$
with $\overline{x}$ the expectancy value of $x$. Respectively, we get 
$
Var({\rm k} \cdot x) = {\rm k}^2 \cdot Var(x)
$ with k = constant.
In particular, the mean value of $\overline{x}$ of $N$ random variables $x_1,...,x_N$ is
$
\overline{x} = \frac{1}{N} \cdot \sum_{i=1}^N x_i
$
and hence
	\[
Var(\overline{x}) = Var \left( \frac{1}{N} \cdot \sum_{i=1}^N x_i \right) = \frac{1}{N^2} 
\cdot \sum_{i=1}^N  Var (x_i).
\]
Note that $\overline{x}$ represents a random variable, as well. If all 
variances $Var (x_i)$ are identical the $\sqrt{N}$-rule gives
$
\sigma (\overline{x}) = \sigma (x)/\sqrt{N}.
$

In their approach C\&M develop $W_{\lambda}$ in a 
multi-dimensional expansion 
 \[
W_{\lambda} = \overline{W} + \sum_{j=1}^M \frac{\partial W_{\lambda}}{\partial F_j} 
\cdot (\overline{F_j} - F_j) + \frac{\partial W_{\lambda}}{\partial F_c} \cdot 
(\overline{F_c} - F_c). 
\]
We interpret the C\&M notation of $\overline{F_j}$ and $\overline{F_c}$ as 
follows: 
$
\overline{F_j} = \frac{1}{M} \cdot \sum_{j=1}^M F_i 
$
and 
$
\overline{F_c} = \frac{1}{\tilde{N}} \cdot \sum_{i=1}^{\tilde{N}} F_{c_i}
$
with $\tilde{N}$ the number of pixel fluxes for the determination of the noise. 
Using 
$
\frac{\partial W_{\lambda}}{\partial F_j} = - \frac{M \cdot h_{\lambda}}{\overline{F_c} \cdot M} 
$
they get
	\[
W_{\lambda} = - \frac{M \cdot h_{\lambda}}{\overline{F_c}} \cdot \frac{1}{M} \cdot  
\sum_{j=1}^M (\overline{F_j} - F_j) + \frac{\partial W_{\lambda}}{\partial F_c} \cdot 
(\overline{F_c} - F_c).
\]
By calculating the variance of $W_{\lambda}$ one obtains
	\[
Var(W_{\lambda}) = \left( \frac{M \cdot h_{\lambda}}{\overline{F_c}} \right)^2 \cdot 
Var \left( \frac{1}{M} \cdot  \sum_{j=1}^M (\overline{F_j} - F_j) \right) 
\] 
	\[+ \left( \frac{\partial W_{\lambda}}{\partial F_c} \right)^2 \cdot Var 
	(\overline{F_c} - F_c). 
\] 

In their approach C\&M now assume that both average values $\overline{F}$ and 
$\overline{F_c}$ are constant (although this assumption is just an 
approximation for $\overline{F}$ by definition) and hence one gets
	\[
Var(W_{\lambda}) = \left( \frac{M \cdot h_{\lambda}}{\overline{F_c}} \right)^2 \cdot 
Var \left( \frac{1}{M} \cdot  \sum_{j=1}^M F_j \right) 
\] 
	\[
+  \left( \frac{\partial W_{\lambda}}{\partial F_c} \right)^2 \cdot Var (F_c).  
\] 

Now it is clear that C\&M's goal was to estimate the variance within the first 
summand with the variance over the averages $\overline{F} = \frac{1}{M} \cdot 
\sum_{j=1}^M F_j$. The $\sqrt{N}$-rule gives 
	\[
Var(W_{\lambda}) = \left( \frac{M \cdot h_{\lambda}}{\overline{F_c}} \right)^2 \cdot 
\frac{1}{M} \cdot Var(F_j)   
\] 
	\[
	+ \left( \frac{\Delta\lambda - W_{\lambda}}{\overline{F_c}} \right)^2 \cdot 
	Var (F_c)   
\] 
and hence for the weak-line case
	\[
Var(W_{\lambda}) = \left(\frac{M \cdot h_{\lambda}}{S/N}\right)^2 \cdot 
\left(\frac{\overline{F}}{\overline{F_c}}\right)^2 \cdot \frac{1}{M} 
\] 
	\[
+ \frac{Var(F_c)}{\overline{F_c}^2} \cdot (\Delta\lambda - W_{\lambda})^2.
\] 

\begin{figure}
\resizebox{\hsize}{!}
{\includegraphics{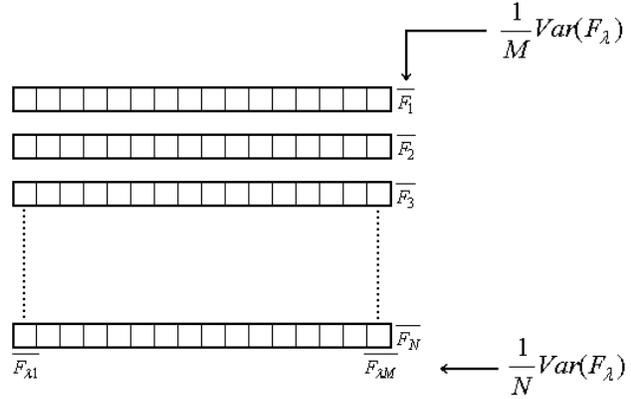}}
\caption{Principle of the variance and average determination for a single 
spectrum and a number of spectra. The figure shows $N$ measured spectra with 
$M$ pixels each.}
\label{label1}
\end{figure}

This is C\&M\,(A9) and the first summand is smaller than our result in 
eq.\,(\ref{eq6}) by the factor $M$. For clarification we show the principal 
procedures in Figure \ref{label1}. 
Every measured single spectrum delivers a separate variance $Var(F_{\lambda})$. 
By obtaining $N$ spectra, the variance for the wavelength-aver-aged spectrum 
is $Var(\overline{F_{\lambda}}) = \frac{1}{N}\cdot Var(F_{\lambda})$. But 
considering the variance of all spectral averages, as is done by C\&M, we get 
$Var(\overline{F}) = \frac{1}{M}\cdot Var(F_{\lambda})$ independent of the 
number of measured spectra. Because of the  missing sum within the continuum 
part of C\&M\,(A3) the factor 1/M does not appear in both sums.    

If we want to estimate $\sigma(W_{\lambda})$ properly by means of C\&M\,(A9), 
depending on the S/N over the whole line, we have to scale up the measured 
noise by the factor $\sqrt{M}$. Therefore the corrected formula C\&M\,(A7) is 
$S/N = \sqrt{M} \cdot \frac{F_j}{\sigma(F_j)}$. 
By using this correction Chalabaev \& Maillard would have found the same result 
for $\sigma(W_{\lambda})$ as we do. 
However, we note that this correction is enforced by the series expansion of 
C\&M.
Their procedure, however, has been used in a series of papers (more than 40 
until now) using expression (A9) of Chalabaev \& Maillard. 

Now it is easy to estimate the correction factor for the hitherto estimated 
error bars of $W_{\lambda}$. For the normal case that both parts $\sigma(F)$ 
and $\sigma(F_c)$ are identical we find
	\[
\sigma(W_{\lambda}) = \sqrt{\frac{2M}{M+1}} \cdot \sigma(W_{\lambda})_{C\&M}.  
\]
For broad lines we have $M+1 \approx M$ and hence 
$
\sigma(W_{\lambda}) \approx \sqrt{2} \cdot \sigma(W_{\lambda})_{C\&M}
$
where the subindex C\&M indicates the values obtained by Chalabaev \& Maillard.
In addition, if we assume that both errors (line and continuum) are different 
(as done by C\&M) our result can differ even more.

In fact, the above eq.\,(\ref{eq10}) for $\sigma^2(W_{\lambda})$ does not 
represent the exact variance of $W$ but is just an approximation, which is 
especially important in the case of small basic populations (i.e. the number 
of pixel fluxes for the determination of noise).
The exact definition of the unbiased variance of $W_{\lambda}$ is
\begin{eqnarray}
\sigma^2(W_{\lambda}) = \frac{\tilde{N} \cdot (M - 1)}{\tilde{N} \cdot M - 1} \cdot 
\left( \frac{\partial W_{\lambda}}{\partial \overline{F}} \cdot \sigma (F) \right)^2 
+\cr 
+\frac{M \cdot (\tilde{N} - 1)}{\tilde{N} \cdot M - 1} \cdot 
\left( \frac{\partial W}{\partial \overline{F_c}} \cdot \sigma (F_c) \right)^2 
\end{eqnarray}
(Kreyszig \cite{kre67}) with $\tilde{N}$ the number of pixel values for the 
estimation of S/N and $M$ the number of pixels within the line. This equation 
transforms into eq.\,(\ref{eq10}) if $\tilde{N}$ and $M$ become sufficiently 
large. If the number of pixels for the estimation of S/N and/or the line width 
are too small, the error of equivalent width will be overestimated by using 
eq.\,(\ref{eq10}). This is especially important for low-resolution observations. For example, if the investigated line covers only 10 pixel and S/N is estimated over about 50-100 pixels, the true error would be overestimated by about 3\%.

\subsection{Scrutinizing previous results}
Former conclusions regarding $W_{\lambda}$ have to be carefully scrutinized due 
to increased error bars with respect to the result of C\&M, especially if the 
interpretation of the data depends on results at the detection limit. 
For clarification, we give two prominent examples:
\begin{itemize}
   \item Searching for rapid line variability in the Be star BD +37°1160, 
   Rossi et al. (\cite{ros91}) tried to find a $W_{\lambda}$ correlation between 
   H$\alpha$ and $H\beta$ (see their Fig.\,4). Their conclusion of 
   "{\it ...spectral line variability on a time scale of ~300 s} for this Be 
   star is unsustainable, especially for $H\beta$.
   \item Investigating periodic variabilities in UV lines of the B star 
   $\zeta$\,Cas, Neiner et al. (\cite{nei03}) claimed to see minima and maxima 
   for $W_{\lambda}$ of the N\,V 1240 doublet (see their Fig.\,2). This is not 
   valid with respect to our correction factor of $\sqrt{2}$. 
\end{itemize}

\begin{acknowledgements}
The authors would like to thank Tony Moffat and G\"unter Gebhard for helpful 
discussions. K.V. would like to thank his wife Susanne for her patience. T.E 
would like to thank Karin and Karl-Werner Eversberg for their support. 
\newline \newline
\textbf{This publication is dedicated to Karl-Werner Eversberg who passed away in October 2005.} 
\end{acknowledgements}


\begin{thebibliography}{}

\bibitem[1983]{cha83}
Chalabaev, A., Maillard, J.P.: 1983, A\&A  127,  279 

\bibitem[1967]{kre67}
Kreyszig, E.: 1967, {\it Statistische Methoden und ihre Anwendungen}, 
G\"ottingen; Vandenhoeck \& Ruprecht, p.333 

\bibitem[1977]{lac77}
Lacy, D.: 1977,  ApJ  212,  132 

\bibitem[2003]{nei03}
Neiner, C., Geers, V.C., Henrichs, H.F., Floquet, M., 
Fr$\acute{{\rm e}}$mat, Y., Hubert, A.-M., Preuss, O., Wiersema, K.:  2003,  
A\&A  406,  1019

\bibitem[1991]{ros91}
Rossi, C., Norci, L., Polcaro, V.F.: 1991, ApJL  249,  L19

\bibitem[1974]{wil74}
Williams, W.E., Frantz, R.L., Breger, M.: 1974, A\&A  35,  381 

\bibitem[1974]{you74}
Young, A.G.: 1974, in:  N. Carleton (ed.), {\it Methods of 
Experimental Physics: Astrophysics},  New York, Academic Press, p.101 

\end{thebibliography}
\end{document}